\documentclass[a4paper]{article}
\usepackage{amsfonts}
\usepackage{amssymb}
\usepackage[margin=2cm]{geometry}
\usepackage[dvips]{graphicx}

\usepackage[T1]{fontenc}
\usepackage[latin2]{inputenc}

\usepackage{amsmath}
\usepackage{bbm}

\begin{document}


\title{Quantum decay cannot be completely reversed. The $5\%$ rule. }

\author{Robert Alicki \\ 
  {\small
Institute of Theoretical Physics and Astrophysics, University
of Gda\'nsk,  Wita Stwosza 57, PL 80-952 Gda\'nsk, Poland}\\
}

\date{\today}
\maketitle

\begin{abstract}
Using an exactly solvable model of the Wigner-Weisskopf atom it is shown that an unstable quantum state cannot be recovered completely by the procedure involving detection of the decay products followed by creation of the time reversed decay products state, as proposed in \cite{Son}. The universal lower bound on the recovery error is approximately equal to $5\% $ of the \emph{error per cycle} - the dimensionless parameter characterizing decay process in the Markovian approximation. This result has consequences for the efficiency of quantum error correction procedures which are based on syndrome measurements and corrective operations.

\end{abstract}

One of the most discussed problems in physics is the origin of macroscopic irreversibility despite the microscopic reversibility of (almost) all known laws of Physics. The standard explanation relies on the unavoidable loss of information about correlations between microscopic constituents of a macroscopic system. On the other hand, it seems, that for a small quantum system  there are no fundamental obstacles to recreate its initial state with an arbitrarily high fidelity. Consider, as an example,  the spontaneous emission from a two-level atom at zero temperature which is a paradigmatic irreversible process. The first method of the initial state recovery for the excited atom in vacuum is to put the atom into an optical cavity. Then due to Poincar\'e recurrences the emitted photon is reabsorbed and one observes a sequence of revivals of the initial state \cite{Brune}. Obviously, the revivals are not perfect, because the cavity is not an ideally isolated system and dissipates energy to the external world. The ultimate presence of the external world can be always modeled by considering the spontaneous emission process in an infinite space. 
In this setting one can again try to recover the initial state by performing a measurement which detects the emitted photon and then  sending a properly designed single-photon state which corresponds to the time-reversed emitted wave-packet. An ingenious experiment realizing this idea has been proposed in \cite{Son}. The question arises whether in principle
and under ideal conditions the initial state can be recovered with the fidelity arbitrarily close to one. I am going to show that this is not the case.

Consider a model of the Wigner-Weisskopf atom (for the rigorous analysis of this model, see \cite{Jak}) with the Hilbert space spanned by a single vector $|e\rangle$ corresponding to the initial excited state of the atom + the vacuum state of the field and the manifold of single photon states (wave packets)
 $\{ |\phi\rangle\}$ representing the ground state of the atom + emitted photon. The final result does not depend on the detailed structure of the decay product's (photon's) Hilbert space, only continuity of its energy spectrum matters. Hence, for brevity,  I can treat the wave packet as a function of the angular frequency (or energy)
only, $\phi(\omega), \omega \geq 0$, and the Hamiltonian can be written as ($\hbar \equiv 1$)
\begin{equation}
H = H_0 + V\ ,\ H_0 |e\rangle= \omega_0 |e\rangle \ ,\ H_0\phi(\omega)= \omega \phi(\omega)\ ,\ 
V = |e\rangle\langle g| + |g\rangle\langle e|
\label{ham1}
\end{equation} 
with the \emph{form-factor} $g(\omega)$ describing the localized coupling of the atom to the field. After detection of the emitted photon followed by the creation of the designed wave packet ${\tilde\psi}$, say at time $t_0 = 0$, the system evolves according to the full Hamiltonian $H$ yielding after time $t$ the state
\begin{equation}
|\Psi(t)\rangle=e^{-iHt}|{\tilde\psi}\rangle = e^{-iHt}e^{iH_0t}|\psi\rangle
\label{state}
\end{equation} 
which should be as close as possible to the initial state $|e\rangle$. Here the wave packet ${\tilde\psi}$ has been written in the form $e^{iH_0t}\psi$ . This is always mathematically possible and has a physical meaning as for longer arrival times $t$ one needs wave packets created far away from the atom. This is achieved by shifting "back in time" by the free dynamics the packet $\psi$ localized in the neighborhood of the atom. The continuous spectrum of a photon in an infinite space and the localized character of the atom-photon interaction implies the convergence 
\begin{equation}
\lim_{t\to\infty} e^{-iHt}e^{iH_0t}|\psi\rangle = W|\psi\rangle
\label{wave}
\end{equation} 
where $W$ is the M\o ller wave operator.  The convergence is fast on the time  scale corresponding to the \emph{scattering time} and therefore in the considered situation one can replace $W(t)= e^{-iHt}e^{iH_0t}$ by $W$.
Using the identity
\begin{equation}
W(t) =  1 -i \int_0^t W(s)\bigl(e^{-i\omega_0 s}|e\rangle\langle g_s|+ h.c.\bigr)ds
\label{inteq}
\end{equation} 
one can compute  the  probability amplitude for the recovery process with the initial packet $\psi$
\begin{equation}
R(\psi) = \langle e|W|\psi\rangle =  -i \int_0^{\infty} \langle e| e^{-iHs} |e\rangle \langle g_s|\psi\rangle ds = \langle \phi_0|\psi\rangle
\label{pamp}
\end{equation} 
where $\phi_0$ is  given by
\begin{equation}
\phi_0(\omega)= i\Bigl( \int_0^{\infty} e^{i\omega t}\langle e| e^{-iHt} |e\rangle dt\Bigr) g(\omega) = i{\tilde S}(-i\omega +0)g(\omega)\ .
\label{em}
\end{equation} 
The Laplace transform ${\tilde S}(z)$ of the survival amplitude $S(t)=\langle e| e^{-iHt} |e\rangle$  can be easily computed. Introducing two functions $F(t) = \langle e|W(t)|e\rangle = S(t)e^{-i\omega_0 t}$ and $G(t) = \langle e|W(t)|g_t\rangle $ one obtains from (\ref{inteq}) the coupled equations
\begin{equation}
F(t)= 1- i\int_0^t  G(s)e^{i\omega_0 s}ds\ ,\ G(t)= - i\int_0^t  F(s)e^{-i\omega_0 s} M(t-s)ds
\label{eqa}
\end{equation}
with $M(t) = \langle g|g_t\rangle$. The equations (\ref{eqa}) are solved by the Laplace transform and yield
\begin{equation}
{\tilde F}(z)= \frac{1}{z + {\tilde M}(z-i\omega_0)}\ .
\label{eqa1}
\end{equation}
This allows to compute ${\tilde S}(-i\omega+0)=[i(\omega -\omega_0) + \gamma(\omega)]^{-1}$ with the frequency dependent decay rate
\begin{equation}
\gamma (\omega) = \pi |g(\omega)|^2\ .
\label{dec}
\end{equation} 
A standard renormalization of the frequency $\omega_0$ has been also performed. The  fidelity of the recovery process ${\cal F}(\psi) = |R(\psi)|^2$ is maximal for
the choice $\psi = \phi_0/\|\phi_0\|$ and is given by the \emph{exact expression}
\begin{equation}
{\cal F}_{max} = \|\phi_0\|^2= \frac{1}{\pi}\int_0^{\infty}\frac{\gamma(\omega) d\omega}{(\omega - \omega_0)^2 + \gamma(\omega)^2}< 1\ .
\label{fid1}
\end{equation}
In the weak coupling Markovian approximation one can replace in (\ref{fid1}) $\gamma(\omega)$ by $\gamma\equiv\gamma(\omega_0)<< \omega_0$
to obtain (\ref{fid})
\begin{equation}
{\cal F}_{max} =  \frac{1}{\pi}\int_0^{\infty}\frac{\gamma d\omega}{(\omega - \omega_0)^2 + \gamma^2}
\simeq 1 -\frac{1}{\pi}\int_{-\infty}^{0}\frac{\gamma d\omega}{(\omega - \omega_0)^2 }= 1- \frac{1}{\pi}\frac{\gamma}{\omega_0}\ .
\label{fid}
\end{equation}

The presented model, although simplified, captures all essential features of the decay process into open space, or in other words
with decay products having continuous energy spectrum. The exact formula (\ref{fid1}) and its Markovian approximation (\ref{fid}) are universal, at least in the leading Born approximation which happens to be exact for the Wigner-Weisskopf
model. An even more universal form can be obtained introducing two dimensionless quantities: the \emph{error per cycle} 
given by $\eta =\gamma\tau = 2\pi \gamma/\omega_0$, $\omega_0 = 2\pi/\tau$  and the \emph{minimal error of recovery} $\epsilon_{min} = 1- F_{max}$. Hence (\ref{fid}) is equivalent to
\begin{equation}
\epsilon_{min}= \frac{1}{2\pi^2}\eta \simeq 0.05 \eta\ .
\label{er}
\end{equation}
Another relation can be obtained for the scheme of many measurement and correction cycles. To preserve the initial excited state
during the time $t$ one needs, on the average,  $n = \gamma t$ measurements followed by corrective operations. Then the fidelity of the initial state under perfect conditions (perfect measurement and wave packet preparation) is given by
\begin{equation}
{\cal F}(t) = \Bigl[1- \frac{1}{\pi}\frac{\gamma}{\omega_0}\Bigr]^n \simeq e^{-\eta_{corr} t/\tau}\ .
\label{fid2}
\end{equation}
where 
\begin{equation}
\eta_{corr} \simeq 0.05 \eta^2
\label{eta}
\end{equation}
can be called a \emph{corrected error per cycle}.
\par
Heuristically, the bound (\ref{er}) can be seen as a manifestation of the following thesis:

\emph{An unstable quantum state cannot be prepared with the probability equal to 1, the deviation from 1 is always of the order
of
\begin{equation}
 \frac{\hbar \Gamma} {E}
\label{dev}
\end{equation}
where $E$ is the energy scale used to separate the unstable state from the other states of the system and $\Gamma$ is the decay rate of this state.}
\par
The above statement is a simple consequence of the Heisenberg relation for energy and time. If a given state differs from the others by the energy $E$ we need at least  time $T$ of the order $\hbar/E$ to perform the measurement which can separate this state. The same relation $ET\simeq \hbar$ holds for the time $T$ needed to "rotate" a system from a known stable  state (ground state) to an orthogonal unstable one using the energy level splitting $E$ \cite{Margo}. In both cases, due to irreversible processes the loss of fidelity during time $T$ is of the order of $\Gamma T\simeq {\hbar \Gamma/E}$.
\par
For physical realizations of the spontaneous emission process the values of (\ref{er}), (\ref{eta}) are too small to be reached
experimentally, for instance, using the ideas of \cite{Son}. However,
the bounds (\ref{er}), (\ref{eta}) have some fundamental meaning for the theory of quantum error correction, an important issue in quantum information processing \cite{Niel}.
The active error correction relies on the measurement of a \emph{syndrome} which determines whether a qubit has been corrupted by noise. Then the error  is reversed by applying the \emph{corrective operation} based on the syndrome.
For the discussed model of the state recovery the photon's detection corresponds to the measurement of a syndrome and creation of the optimally designed single-photon wave packet is the corrective operation. The parameter $\eta$ is called an \emph{error per gate} in the context of quantum information processing. As the presented model is paradigmatic for quantum irreversible processes and, moreover, unlimited resources like perfect measurement and perfect state preparation are used  
in the recovery process, I claim that the \emph{$5\%$ rules} (\ref{er}),(\ref{eta}) provide  relevant bounds on the efficiency of any active measurement based error correction procedure. In order to compare this result with those based on error correcting codes and active error correction the distinction between \emph{error correction} and \emph{error prevention} should be made. Any error correction scheme can be described in terms of subsystems \cite{Knill}, such that the protected encoded qubit corresponds to a certain 2-level subsystem and the other degrees of freedom can be treated as a part of an environment. To obtain an arbitrarily high level of protection we have to use  \emph{self-correcting} systems which reduce the effective coupling
of the encoded qubit to environment fast enough with the increasing size of the code. It follows from the fact that, as shown above, the active part of an error correction scheme gives only a universally bounded reduction of error independent of the details
of correction procedures. This provides another argument that the existence or nonexistence of self-correcting quantum memory
is a fundamental question for the feasibility of fault-tolerant quantum information processing \cite{Den,Ali}.

{\bf Acknowledgments} The author is grateful to M. Stobi\'nska,  G. Leuchs  for discussions on \cite{Son} and M. Horodecki for comments. This work is supported by the Polish research network LFPPI.


\begin{thebibliography}{99}
\bibitem{Son}  N. Lindlein, R. Maiwald, H. Konermann, M. Sondermann, U. Peschel and G. Leuchs, Laser Physics,{\bf 17} (7),  927, (2007)     
\bibitem{Brune} M. Brune, F. Schmidt-Kaler, A. Maali, J. Dreyer, E. Hagley, J. M. Raimond, and S. Haroche, Phys.Rev.Lett. {\bf 76}, 1800, (1996).
\bibitem{Jak} V. Jaksic , E. Kritchevski and C-A. Pillet, in
\emph{Large Coulomb Systems}, LNP 695, Springer, Berlin (2006), 145-215.
\bibitem{Margo}
N. Margolus and L.B. Levitin, Physica {\bf D120}, 188, (1998).
\bibitem{Niel} M.A. Nielsen and I.L. Chuang, \emph{Quantum Computation and Quantum Information}, Cambridge University Press (2000).
\bibitem{Knill} E. Knill, R. Laflamme and L. Viola, Phys.Rev.Lett.{\bf 84} , 2528 , (2000).
\bibitem{Den} E. Dennis, A. Kitaev, A. Landahl and J. Preskill, J.Math.Phys.{\bf 43} , 4452, (2002).
\bibitem{Ali} R. Alicki, M. Fannes and M. Horodecki, J. Phys. A: Math. Theor.{\bf 40}, 6451, (2007)   
\end{thebibliography}
\end{document}